\documentclass[aps,prl,reprint,showpacs,amssymb,amsmath,superscriptaddress,longbibliography]{revtex4-1}

\usepackage[T1]{fontenc}
\usepackage{graphicx}
\usepackage[usenames, dvipsnames]{xcolor}
\usepackage{times}
\usepackage{ulem}
\usepackage{verbatim}
\usepackage{siunitx}
\usepackage{bm}

\usepackage[hidelinks]{hyperref}

\usepackage{tikz}
\usetikzlibrary{shapes}

\newcommand{\mybullet}[2][none]{\raisebox{1pt}{\tikz{\node[draw=#1,scale=0.5,circle,fill=#2](){};}}}
\newcommand{\mytriangle}[3][none]{\raisebox{1pt}{\tikz{\node[draw=#1,scale=0.35,regular polygon, regular polygon sides=3,fill=#2,rotate=#3](){};}}}
\newcommand{\mysquare}[2][none]{\raisebox{1pt}{\tikz{\node[draw=#1,scale=0.5,regular polygon, regular polygon sides=4,fill=#2](){};}}}
\newcommand{\mydiamond}[2][none]{\raisebox{0.5pt}{\tikz{\node[draw=#1,scale=0.4,diamond,fill=#2](){};}}}
\newcommand{\mypentagon}[3][none]{\raisebox{1pt}{\tikz{\node[draw=#1,scale=0.5,regular polygon, regular polygon sides=5,fill=#2,rotate=#3](){};}}}

\definecolor{myblue}{HTML}{4477AA}
\definecolor{red}{HTML}{E13311}
\definecolor{teal}{HTML}{009988}
\definecolor{purple}{HTML}{882255}
\definecolor{grey}{HTML}{646464}

\newcommand{\Rb}{$^{87}$Rb }

\newcommand{\ket}[1]{\lvert#1\rangle}

\newcommand{\comm}[1]{}

\begin{document}
\title{Observation of Atom Number Fluctuations in a Bose-Einstein Condensate}

\author{M.\,A.\,Kristensen}
\author{M. B. Christensen}
\author{M.\,Gajdacz}
\affiliation{Institut for Fysik og Astronomi, Aarhus Universitet, Ny Munkegade 120, 8000 Aarhus C, Denmark.}
\author{M.\,Iglicki}
\affiliation{Center for Theoretical Physics, Polish Academy of Sciences, Al. Lotnik\'{o}w 32/46, 02-668 Warsaw, Poland.}
\affiliation{Faculty of Physics, University of Warsaw, Pasteura 5, PL-02093 Warsaw, Poland.}
\author{K.\,Paw\l{}owski}
\affiliation{Center for Theoretical Physics, Polish Academy of Sciences, Al. Lotnik\'{o}w 32/46, 02-668 Warsaw, Poland.}
\author{C.\,Klempt}
\affiliation{Institut f\"ur Quantenoptik, Leibniz Universit\"at Hannover, Welfengarten 1, 30167 Hannover, Germany.}
\author{J.\,F.\,Sherson}
\affiliation{Institut for Fysik og Astronomi, Aarhus Universitet, Ny Munkegade 120, 8000 Aarhus C, Denmark.}
\author{K.\,Rz\k{a}\.{z}ewski}
\affiliation{Center for Theoretical Physics, Polish Academy of Sciences, Al. Lotnik\'{o}w 32/46, 02-668 Warsaw, Poland.}
\author{A.\,J.\,Hilliard}
\author{J.\,J.\,Arlt}
\affiliation{Institut for Fysik og Astronomi, Aarhus Universitet, Ny Munkegade 120, 8000 Aarhus C, Denmark.}

\begin{abstract}
Fluctuations are a key property of both classical and quantum systems. While the fluctuations are well understood for many quantum systems at zero temperature, the case of an interacting quantum system at finite temperature still poses numerous challenges. Despite intense theoretical investigations of atom number fluctuations in Bose-Einstein condensates, their amplitude in experimentally relevant interacting systems is still not fully understood. Moreover, technical limitations have prevented their experimental investigation to date.
Here we report the observation of these fluctuations. Our experiments are based on a stabilization technique, which allows for the preparation of ultracold thermal clouds at the shot noise level, thereby eliminating numerous technical noise sources. Furthermore, we make use of the correlations established by the evaporative cooling process to precisely determine the fluctuations and the sample temperature.  This allows us to observe a telltale signature: the sudden increase in fluctuations of the condensate atom number close to the critical temperature.
\end{abstract}

\maketitle
The experimental realization of weakly interacting Bose-Einstein condensates (BECs) created a new experimental paradigm for understanding many particle quantum systems~\cite{Bloch2008}.
However, the population statistics of BECs has not been investigated experimentally beyond its first moment, since all higher moments, such as the BEC number fluctuations, were inaccessible due to technical noise.
Furthermore, the theoretical description of these higher moments of weakly interacting trapped gases poses considerable challenges~\cite{Kocharovsky2006}, and to date no exact prediction is available at the typical experimental atom numbers.

Historically, the theory of Bose-Einstein condensation for noninteracting bosons was developed within the grand canonical ensemble~\cite{Dalfovo1999}. 
However, the grand canonical description is inadequate to describe atomic Bose gases of fixed atom numbers below the critical temperature for condensation~\cite{Ziff1977}. In particular, as the ground state becomes macroscopically occupied, grand canonical theory predicts unphysically large fluctuations of the total atom number $ \Delta N \sim N $, which contradicts particle conservation. This observation is at times referred to as the grand canonical catastrophe~\cite{Grossmann1996,Holthaus1998}. Any description of the system must therefore be based on the canonical or microcanonical ensemble.

\begin{figure}[tb]
	\centering
	\includegraphics[width=1\linewidth]{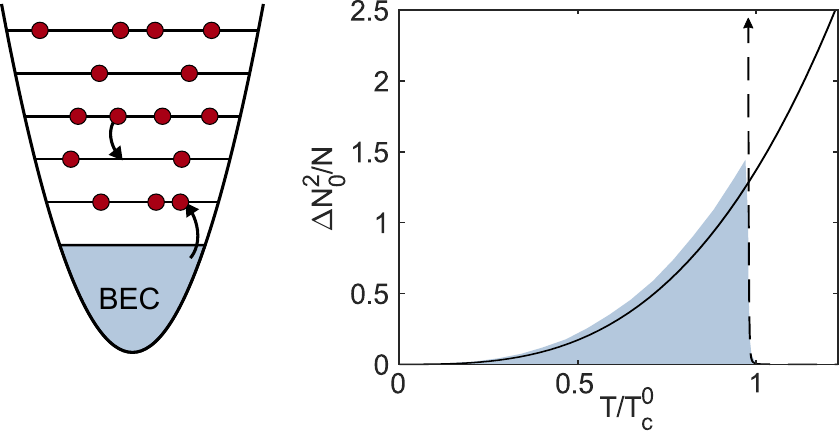}
	\caption{
	Fluctuations of a noninteracting BEC in an isotropic harmonic trap. Left: Illustration of the fluctuation process due to continuous particle exchange between energy levels. Right: When the temperature is lowered towards the critical temperature $T_\mathrm{c}^0$ a grand canonical ensemble calculation predicts unphysically large fluctuations below $T_\mathrm{c}^0$ the (dashed line). 
	Below the critical temperature, the leading contribution Eq.~(\ref{eq:Politzer}) provides a good estimate of the fluctuations~\cite{Politzer1996} (solid line). An exact numerical calculation shows how these two expressions serve as limiting cases in the two temperature regions~(shaded area).}
	\label{fig:intution}
\end{figure}

Renewed theoretical interest in the number fluctuations in a BEC was inspired by the experimental realization of Bose-Einstein condensation in dilute gases~\cite{Ketterle1999}.
The asymptotic expression for the variance of the condensate population in a canonical ensemble was first derived by Politzer~\cite{Politzer1996} for an ideal gas
\begin{equation}\label{eq:Politzer}
\Delta N_0^2 = \frac{\zeta(2)}{\zeta(3)}N\left(\frac{T}{T_\mathrm{c}^0}\right)^3,
\end{equation}
where $ T_\mathrm{c}^0 $ is the ideal gas critical temperature for Bose-Einstein condensation and $ \zeta(x) $ is the Riemann zeta function.
This was quickly followed by the introduction of the Maxwell demon ensemble~\cite{Navez1997} that enabled the derivation of the asymptotic expression in the microcanonical ensemble. 
In parallel, numerical techniques were introduced for exact calculations for a finite number of particles~\cite{Weiss1997}, and a master equation approach based on the laser phase transition analogy was developed~\cite{Scully1999}.
Figure~\ref{fig:intution}  illustrates the breakdown of the grand canonical ensemble and the behavior of the fluctuations for an ideal Bose gas.

In general, an ensemble description is physically meaningful only in the presence of interactions.
Despite numerous attempts~\cite{Giorgini1998, Idziaszek1999, Kocharovsky2000, Kocharovsky2006, Svidzinsky2006,Bhattacharyya2016}, no exact results are known for interacting bosons, especially close to the critical temperature. 
An experimental investigation of this fundamental feature is therefore of paramount importance.
In recent experiments fluctuations in photon BECs were studied under grand canonical conditions where the particle number conservation does not apply~\cite{Schmitt2014,Wurff2014}. 
However, for atomic BECs only the statistics of the total number of particles has been studied \cite{Chuu2005}.

In this letter, we observe the fundamental fluctuations of the ground state occupation of harmonically trapped interacting BECs. 
Despite the considerable theoretical attention outlined above, the fluctuations have to date eluded experimental observation because fluctuations on the order of $ \sqrt{N} $ are typically hidden in the technical fluctuations of BEC experiments.
We have recently implemented a stabilization technique that permits the preparation of ultracold thermal clouds at the shot noise level~\cite{Gajdacz2016},
which eliminates numerous technical noise sources. 
A final evaporative cooling step from this starting point thus allows for the production of BECs in a precisely controlled range of atom numbers.
We find that both the BEC occupation and the temperature are strongly correlated with the total atom number. 
Based on this correlation, both the fluctuations and 
the temperature can be precisely determined, thereby allowing the observation of BEC fluctuations.

The experimental challenges for the observation of fluctuations are illustrated in Fig.~\ref{fig:intution}. The variance of the peak number fluctuations of a BEC  in a typical realization is $\Delta N_0^2 \sim1.5 N $. For a total atom number of $N\sim 5\times10^5$, this implies that  the standard deviation of the BEC atom number $N_0 $ must be measured at a level of $\Delta N_0\approx 700$. 
Thus, the relative precision of the atom number determination must be better than $ \frac{\Delta N}{N} < \SI{0.14}{\percent} $
\footnote{In principle this can also be achieved in a system with large shot-to-shot fluctuations. If $N$, $N_0$ and $T$ can be determined sufficiently well, the data can be binned according to $N$ and $T$ to obtain $\Delta N_0$. In practice however, the determination of $T$ from individual fits to the experimental data is typically not sufficiently precise. Moreover this approach would require a very large number of realizations, due to the variation of $N$ and $N_0$ by $10\%$ in typical experiments}.
In the following, we describe how this precision is achieved in our experimental realization.

The experimental apparatus used to produce BECs has been described in detail previously~\cite{Park2012}. Briefly,  \Rb atoms are initially captured in a magneto-optical trap and transported into a Ioffe-Pritchard-type magnetic trap where they are cooled by radio-frequency (RF) evaporation. To stabilize the production of ultracold clouds, the cooling process is interrupted when the clouds contain $\sim 4 \times 10^6$ atoms at a temperature of $\sim 14$~$\mu$K.
The stabilization technique was presented in~\cite{Gajdacz2016} and here we describe only the improvements to the procedure relevant for the current experiment.
We probe the clouds using minimally destructive Faraday imaging \cite{Gajdacz2013, *Kristensen2017} by acquiring 50 images that are analyzed in real time. 
Based on the outcome, the atom number is corrected by spilling excess atoms using a weak RF-pulse of controllable duration. The RF-pulse is resonant with atoms at the mean energy of the sample and thus does not affect the temperature of the sample.  
Subsequently, the magnetic trap is decompressed by increasing the bias field in the axial direction leading to radial and axial trapping frequencies of  $\omega_\rho = 2\pi \times \SI{93.4}{\hertz} $ and $\omega_z = 2\pi \times \SI{17.7}{ \hertz}$~
\footnote{The decompression lowers the trap anisotropy and thus reduces the probability for the excitation of phase fluctuations. Thus the occurrence of the density modulations after time-of-flight expansion is avoided, which can be detrimental for the evaluation of atom numbers and temperatures.}  and an aspect ratio of $ 5.27 $.
At this point, the clouds are probed by a second set of 20 Faraday images to ensure that the stabilization was successful. Finally, BECs are produced by applying a last RF sweep that ends at a frequency corresponding to the desired average BEC occupation.
To ensure that the clouds are in thermal equilibrium, they are first held in the trap for a further 3 seconds at the final radio frequency and \SI{1}{\second} without RF before the trap is extinguished.
Finally, the clouds are probed using resonant absorption imaging after a \SI{27.5}{\milli\second} time of flight.

To accurately detect the density distribution of dense clouds we use saturating light at an intensity $I=2.3~I_\mathrm{sat}$, and calibrate the imaging system following Ref.~\cite{Reinaudi2007}. 
Off-resonant light in the imaging beam limits the maximal observable optical density and must be minimized. This is achieved by using a volume Bragg grating, which removes the broad background of spontaneous emission typically emitted by diode lasers. 
Vibrations of optical components in the imaging system degrade the ability to normalize the absorption image with a background image, and thus deteriorate the result of absorption imaging.
To mitigate this effect, the separation between the images is minimized. By applying an optical pumping pulse on the $ \ket{F = 2} $ to $ \ket{F' = 2} $ transition after the absorption image, all atoms are transferred to the $ \ket{F = 1} $ state. The cloud thus becomes transparent to the background imaging pulse and allows us to reduce the image separation to \SI{340}{\micro\second}, which is limited by the camera readout speed.

The atom number and temperature are extracted from the time-of-flight images as follows. The optical density is integrated by summation in a region of interest (ROI) containing the entire cloud to obtain the total atom number $N$. Then the optical density is fitted with a Bose-enhanced thermal distribution in a ring-shaped ROI, which excludes the condensate, to obtain the temperature $T$ of the sample. This thermal distribution is subtracted from the total optical density, and the condensate atom number $N_0$ is obtained by summation of the remaining optical density in the central ROI.
The high quality of the absorption imaging thus allows for a precise determination of the atom numbers $N_0$ and $N$. However, the fitting procedure typically results in a larger uncertainty of the temperature, which prevents a direct evaluation of the fluctuations $\Delta N_0^2$ at a given temperature. 

\begin{figure}[tb]
	\centering
	\includegraphics[width=1\linewidth]{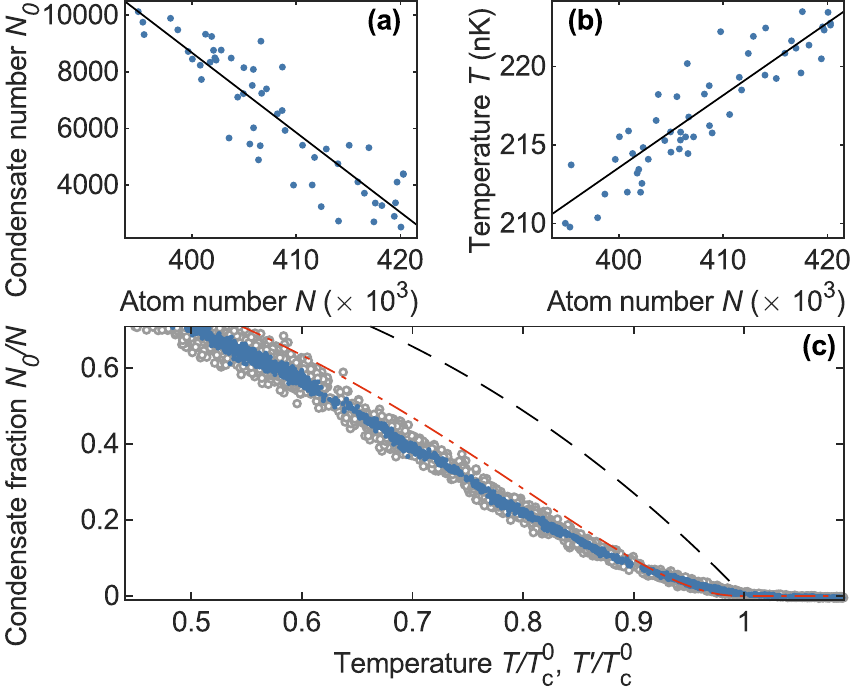}
	\caption{Condensate atom number and temperature as a function of the total atom number~\cite{SupplementaryMaterial}. %~\cite{SupplementaryMaterial,Grossmann1995,Ketterle1996a,Holzmann2004,Giorgini1996,Giorgini1997,Grueter1997,Arnold2001,Kashurnikov2001,Smith2011,Gerbier2004,Bienias2011, Witkowska2009,Karpiuk2012,Pawlowski2013,Witkowska2011,Brewczyk2007}.
	(a) BEC atom numbers and  (b) temperatures obtained from a fit to the optical density. The solid lines are linear fits to the data that are used to obtain the BEC atom number variance  $\Delta N_0^2$ and the corrected temperature $T'$. 
	(c) Condensate fraction $N_0/N$ as a function of the reduced temperature $ T/T_\mathrm{c}^0 $ (open circles) and corrected temperature $ T'/T_\mathrm{c}^0 $ (blue points) obtained from the correlations illustrated in (b). $ T_\mathrm{c}^0 $ is the ideal gas critical temperature. The dashed line represents the ideal gas result in the thermodynamic limit, whereas the dot-dashed red line represents the semi-ideal model including interactions~\cite{Naraschewski1998}. 
	}
	\label{fig:tvsnvsn0}
\end{figure}

This difficulty is overcome due to the stabilization technique in combination with the inherent properties of evaporative cooling. When an evaporative cooling step is applied to an atomic cloud of well-known initial atom number and temperature, the final atom number and temperature are strongly correlated~\cite{Ketterle1996}. In typical experiments the initial atom number is not well controlled, and this correlation is washed out. In our case, however, the stabilization provides a well-known initial atom number and after evaporation $T$ and $N_0$ can be regarded as a function of the total atom number, reducing the problem dimensionality to a single parameter $N$.

Figures~\ref{fig:tvsnvsn0}(a) and \ref{fig:tvsnvsn0}(b) show the BEC atom number and temperature as a function of the total atom number for a chosen final radio frequency. The remaining variation of the total atom number is primarily due to small drifts of the trap's magnetic offset  field, causing minor variations of $N_0$ and $T$. Note that this corresponds to a narrow interval in $ N $, $ N_0 $ and $ T $.
To extract the variance of the BEC number in such an interval we linearly fit $N_0$~\footnote{A linear fit is the most conservative choice on a small interval and avoids over interpretation of the data.} and obtain $\Delta N_0$  by evaluating the variance with respect to the fit. 

In addition, the correlation between $ N $ and $ T $  over the entire data set allows us to extract a precise temperature for each realization. Similar to the BEC atom number, we linearly fit $ T $ as shown in Fig.~\ref{fig:tvsnvsn0}(b). Based on this fit we obtain a corrected temperature $ T' $ for each realization. This method reduces the uncertainty of individual temperatures by using all temperature information within a narrow interval and by using the high precision of atom number determination to obtain the best estimate for the temperature in a particular realization.

Figure~\ref{fig:tvsnvsn0}(c) shows the condensate fraction $ N_0/N $ as a function of the corrected temperature $ T'/T_\mathrm{c}^0 $
where the ideal gas critical temperature $ T_\mathrm{c}^0 $ was calculated individually for each point. The data clearly illustrate the high precision obtained with this method.
The deviation from the ideal gas prediction is primarily caused by repulsive interactions~\cite{Giorgini1996}. The semi-ideal model captures the smooth onset of condensation and the observed condensate fraction very well. The remaining disagreement is due to the noninteracting thermal approximation of the model and systematic errors of the temperature determination~\cite{Gerbier2004,SupplementaryMaterial}.

\begin{figure}[tb]
	\centering
	\includegraphics[width=1\linewidth]{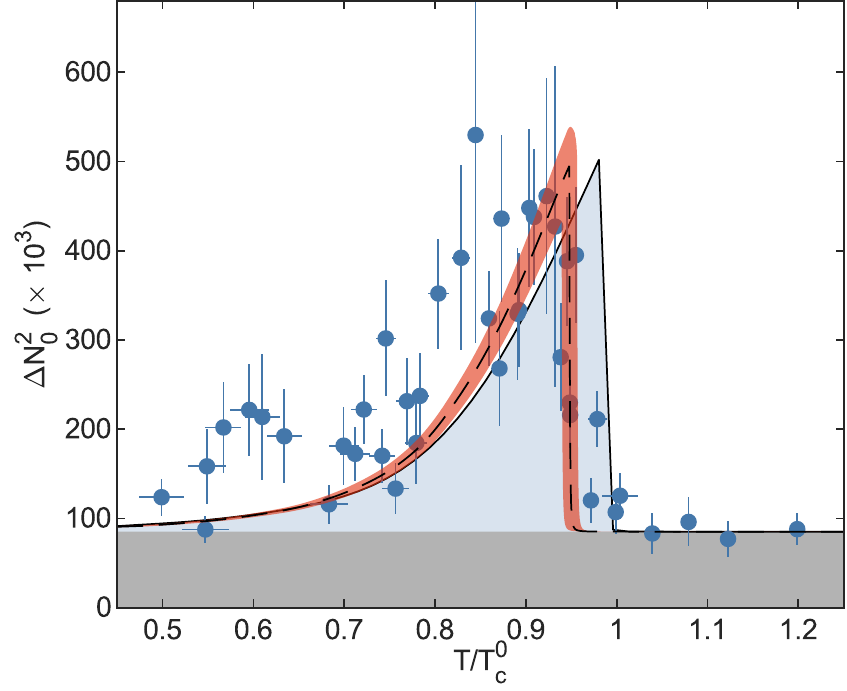}
	\caption{Fluctuations of the BEC. Blue points: Observed variance of the BEC atom number. Vertical error bars are statistical uncertainties of the variance. Horizontal error bars are $ 1\sigma $ spread in the reduced temperature. Dashed line: Fit to the data using the noninteracting approximations (see text). Red shading indicates a $ 1\sigma $ confidence bound on the fit and the step size in $ T/T_\mathrm{c}^0 $ at the onset of fluctuations. Light blue shading: Prediction for the fluctuations of a noninteracting gas using exact particle number counting statistics. Dark gray shading: Offset due to technical fluctuations.}
	\label{fig:fluctuations}
\end{figure}

Figure~\ref{fig:fluctuations} shows our main result, the variance of the condensate number $ \Delta N_0^2 $ as a function of the reduced temperature $T/T_\mathrm{c}^0$, where $ T $ is the mean temperature.
Just below $ T_\mathrm{c}^0 $, the data show a clear onset of fluctuations, as expected from Fig.~\ref{fig:intution}. The observed fluctuations grow rapidly as the critical temperature is crossed, and well below $ T_\mathrm{c}^0 $ they decay due to decreasing $ T $ and $ N $.
To verify this result the experiment was conducted for different  trap aspect ratios and rethermalization conditions~\cite{SupplementaryMaterial}. Additionally, different fitting models of the absorption images were tested~\cite{SupplementaryMaterial}.

Each data point in Fig.~\ref{fig:fluctuations} is based on at least 45 experimental realizations. The number varies slightly, since a small number of  realizations where the stabilization failed were excluded. 
Above $ T_\mathrm{c}^0$ the data are evaluated with the exact same method despite the absence of a BEC. The observed variance therefore corresponds to an offset arising from technical fluctuations. Based on the average variance of the data points at $ T/T_\mathrm{c}^0 > 1 $ we obtain an offset variance of $ \mathcal{O} = 0.78\times10^5 $.
This confirms that atom number variations at the level of $ \sim 300 $ atoms can be detected.

Because of the theoretical challenges posed by this interacting quantum system, a full comparison with a theoretical prediction is not possible at the atom numbers in our experiments. However, in the following we estimate the most relevant contributions beyond the canonical solution due to the geometry of the trap and the effect of interactions.

\begin{figure}[tb]
	\includegraphics[width=1\linewidth]{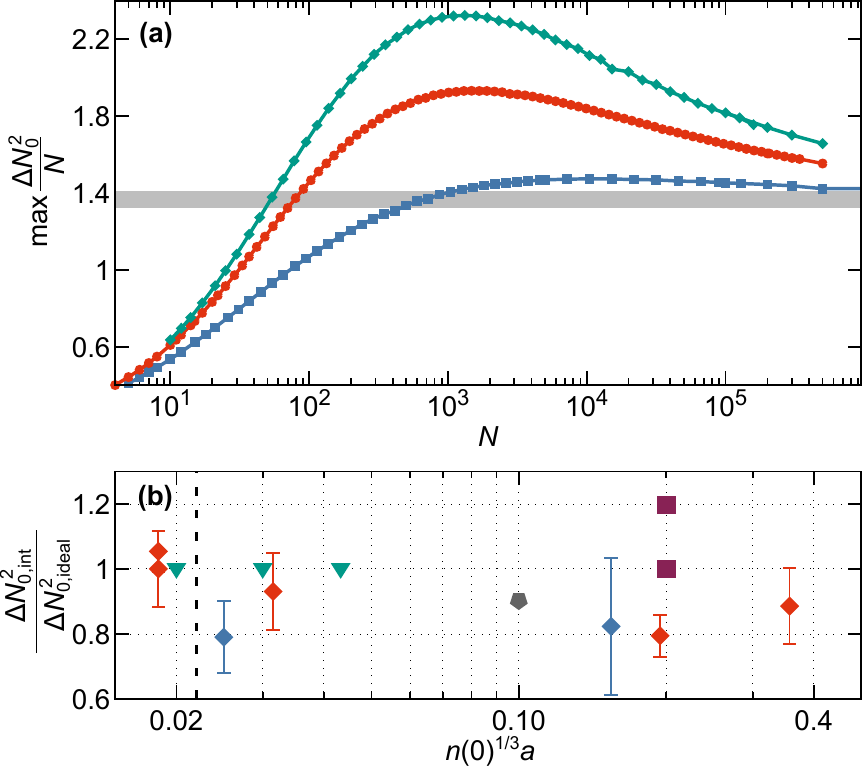}
	\caption{Influence of geometry and interactions on the condensate fluctuations. 
	(a) Maximal fluctuations of the number of condensed particles $\max( \Delta N_0^2 / N)$ as a function of the total number of atoms for the spherical trap (\mysquare{myblue}) and the traps in which experiments have been performed with aspect ratios  $ 5.27 $ (\mybullet{red}) and $ 7.64 $ (\mydiamond{teal})~\cite{SupplementaryMaterial}. The horizontal gray line is the limiting value $\zeta (2) / \zeta (3)$ given in Eq.~(\ref{eq:Politzer}). 
	(b) Fluctuations of the condensate atom number in the interacting system relative to the  noninteracting system as a function of the interaction strength $n(0)^{1/3} a$, where $n(0)$ is the peak density and $ a $ the scattering length. (\mydiamond{myblue}) and (\mydiamond{red}) are results obtained within the classical field approximation for $N=100$ and $N=200$ atoms~\cite{SupplementaryMaterial}. The other points correspond to \mytriangle{teal}{60} \cite{Bhattacharyya2016}, \mysquare{purple} \cite{Idziaszek1999}, \mypentagon{grey}{36} \cite{Svidzinsky2006}.  The vertical dashed line indicates a typical interaction strength in our experiments.
	}
	\label{fig:theoryComparison}
\end{figure}

First the dependence of the fluctuations on the geometry of the trapping potential is analyzed. We calculated the fluctuations for a noninteracting Bose gas in the canonical ensemble using exact particle number counting statistics~\cite{Weiss1997}. Figure~\ref{fig:theoryComparison}(a) shows the maximum relative variance of the fluctuations for a spherical trap and for our elongated trapping geometry as a function of the total atom number. 
For large atom numbers, the limiting value corresponds to the result of Eq.~(\ref{eq:Politzer}) at $T_\mathrm{c}^0$ which yields $\Delta N_{0,\mathrm{max}}^2/N=\zeta(2)/\zeta(3)\approx1.36$. We attribute the behavior in the elongated trap at intermediate atom numbers to the reduced dimensionality of the system at these parameters, which leads to increased fluctuations. The asymmetry of the trapping potential thus plays a crucial role for the expected maximal fluctuations. For our case these  fluctuations are enhanced by 15\% compared to a spherical trap.  

Secondly we discuss the effect of interactions on the fluctuations. Despite considerable theoretical effort~\cite{Kocharovsky2006}, there is currently no model offering an exact prediction for a trapped, interacting BEC for large number of atoms close to the phase transition. We therefore estimate the effect of interaction using a classical fields approach (CFA) for smaller samples of up to $400$ atoms, but adjust the interaction strength to mimic the experiment. 
The CFA has been described in detail in~\cite{Brewczyk2007}. Briefly, we map the quantum field operator $\hat{\Psi} (\bm{r})$ to a classical field decomposed into harmonic oscillator (HO) eigenfunctions of the trap. Thus we map the quantum problem to a classical one described by the complex HO amplitudes. To avoid an ultraviolet catastrophe, we use an appropriate cutoff for the high energy modes~\cite{Witkowska2009}. Finally, we sample the phase space in thermal equilibrium by using the Metropolis algorithm~\cite{Bienias2011,SupplementaryMaterial}. Figure~\ref{fig:theoryComparison}(b) shows our result in terms of the ratio between the maximal fluctuations of an interacting and a noninteracting BEC as a function of the interaction strength $n(0)^{1/3}a$. Additionally, appropriately scaled results from all other available simulations for interacting harmonically trapped BECs are displayed. 
These calculations indicate that the interactions  play only a minor role, and thus a comparison with the ideal gas result is sufficient to analyze the present experimental result.

Based on this understanding we compare our experimental results to noninteracting canonical theories in two ways. First we compare our result with an exact canonical calculation~\cite{Weiss1997}, shown as a blue shaded region in Fig.~\ref{fig:fluctuations} for our trapping geometry and atom number at each $ T/T_\mathrm{c}^0 $. To account for technical fluctuations we have added the constant offset $ \mathcal{O} $ to the theory. The striking increase in fluctuations just below the ideal gas critical temperature is well reproduced in this comparison. The increase is consistent with our data, which shows the same feature, thus demonstrating that we have observed the onset of fluctuations in a BEC. 

Secondly, we compare our results with the limiting noninteracting theoretical cases presented in Fig.~\ref{fig:intution}~\cite{Politzer1996}. Inspired by these results, we choose the fitting function 
\begin{equation} \label{eq:Fit}
\Delta N_0^2 =\min
\begin{cases}
\alpha N \left( \frac{T}{T_\mathrm{c}^0} +\delta t \right)^3 + \mathcal{O} \\[5pt]
N_0(N_0+1) + \mathcal{O},
\end{cases}
\end{equation} 
where the free parameters are the amplitude of the fluctuations, $\alpha$, and the shift of the reduced temperature, $ \delta t$.
The result of the fit is shown in Fig.~\ref{fig:fluctuations} providing $\alpha = 1.65(17)$ and  $ \delta t = -0.040(9)$. The amplitude $ \alpha $ is slightly larger than the limiting value $\alpha = \zeta(2)/\zeta(3) = 1.36 $ for the noninteracting case Eq.~(\ref{eq:Politzer}). This is consistent with an expected enhancement of the fluctuations due to the asymmetry of the trap as shown in  Fig.~\ref{fig:theoryComparison}(a), and hence the agreement with the non-interacting calculation taking trapping geometry into account is very good. 
Additionally, we observe a small shift in the reduced temperature of \SI{4.0}{\percent}, which is expected due to a shift of the critical temperature caused by interactions~\cite{Giorgini1996,SupplementaryMaterial}. The fit also allows us to extract the maximal fluctuations $ \Delta N_{0,\mathrm{max}}^2 = 4.1(8)\times10^5$ at a reduced temperature $ (T/T_\mathrm{c}^0)_\mathrm{max} = 0.96(8) $ and atom number $ N_\mathrm{max} = 2.58(6) \times 10^5 $.

In conclusion, we have observed the atom number fluctuations in a trapped interacting BEC indicated by a telltale increase of the fluctuations at the critical temperature.
The observed fluctuations are well described by a noninteracting theory including the asymmetry of the harmonic trap.
At the present signal-to-noise level, it is not possible to discriminate between different theoretical predictions for the fluctuations in interacting BECs. 
In future experiments we plan to investigate the scaling of the fluctuations with the atom number at fixed temperature, which has been theoretically debated~\cite{Giorgini1998,Idziaszek1999,Yukalov2004,Zwerger2004}.

\begin{acknowledgments}
M.A.K., M.B.C., M.G., J.F.S., A.J.H., and J.J.A. acknowledge support from the  Villum  Foundation,  the  Carlsberg Foundation,  and  the Danish Council for Independent Research. This work was supported by the (Polish) National Science Center Grants No. 2014/13/D/ST2/01883 (K.P.), and No.
2015/19/B/ST2/02820 (K.R. and M.I.). C.K. acknowledges support by the Deutsche Forschungsgemeinschaft (DFG) through CRC 1227 (DQ-mat), project B01.
\end{acknowledgments}

\end{document}

\title{Supplementary material for: Observation of Atom Number Fluctuations in a Bose-Einstein Condensate}

\author{M.\,A.\,Kristensen}
\author{M. B. Christensen}
\author{M.\,Gajdacz}
\affiliation{Institut for Fysik og Astronomi, Aarhus Universitet, Ny Munkegade 120, 8000 Aarhus C, Denmark.}
\author{M.\,Iglicki}
\affiliation{Center for Theoretical Physics, Polish Academy of Sciences, Al. Lotnik\'{o}w 32/46, 02-668 Warsaw, Poland.}
\affiliation{Faculty of Physics, University of Warsaw, Pasteura 5, PL-02093 Warsaw, Poland.}
\author{K.\,Paw\l{}owski}
\affiliation{Center for Theoretical Physics, Polish Academy of Sciences, Al. Lotnik\'{o}w 32/46, 02-668 Warsaw, Poland.}
\author{C.\,Klempt}
\affiliation{Institut f\"ur Quantenoptik, Leibniz Universit\"at Hannover, Welfengarten 1, 30167 Hannover, Germany.}
\author{J.\,F.\,Sherson}
\affiliation{Institut for Fysik og Astronomi, Aarhus Universitet, Ny Munkegade 120, 8000 Aarhus C, Denmark.}
\author{K.\,Rz\k{a}\.{z}ewski}
\affiliation{Center for Theoretical Physics, Polish Academy of Sciences, Al. Lotnik\'{o}w 32/46, 02-668 Warsaw, Poland.}
\author{A.\,J.\,Hilliard}
\author{J.\,J.\,Arlt}
\affiliation{Institut for Fysik og Astronomi, Aarhus Universitet, Ny Munkegade 120, 8000 Aarhus C, Denmark.}

\maketitle

\section{Fluctuations under different experimental conditions}
To assess the influence of trapping geometry and equilibration time the fluctuations of the BEC atom number were investigated under different experimental conditions.
A second data set was obtained for larger atom numbers and a more asymmetric trap with radial and axial trapping frequencies of $\omega_\rho = 2\pi \times 128.5$~Hz and $\omega_z = 2\pi \times 17.7~$Hz.
Additionally, the RF frequency was held at its final frequency for two seconds before trap was turned off, providing different conditions for rethermalization. This data set was used in Fig. 2 of the main text.

\begin{figure}[tbh]
	\centering
	\includegraphics[width=\linewidth]{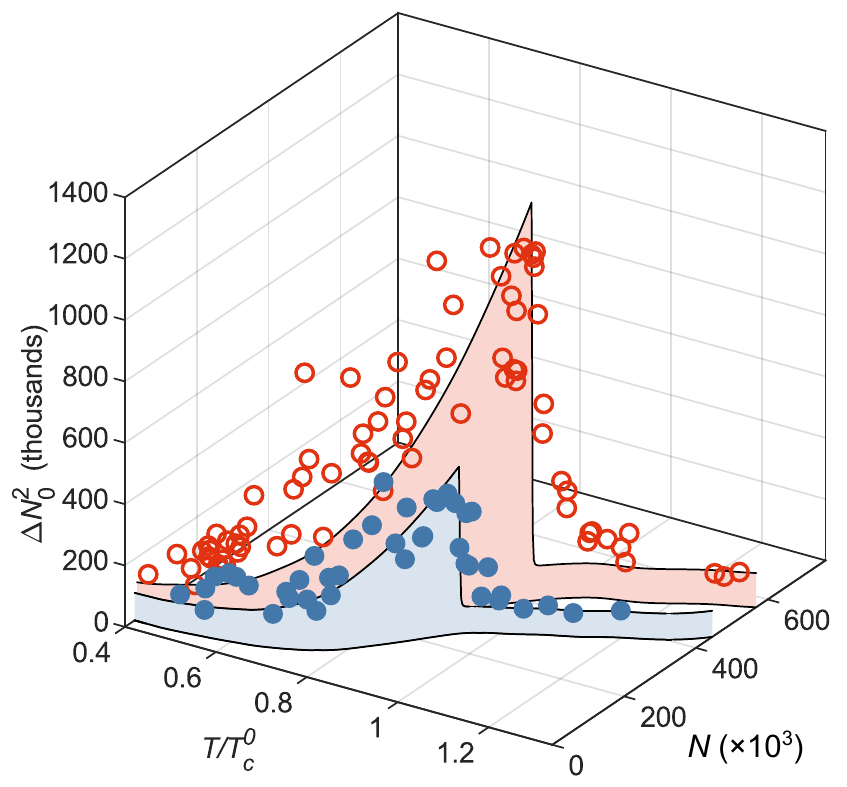}
	\caption{
		Comparison of two measurements of the fluctuations obtained under different experimental conditions. Blue circles correspond to the data presented in the main text. Open red circles is data obtained at larger atom numbers in a more asymmetric trap. The shaded areas are fits of Eq.~(2) in the main text to the data.
	}
	\label{fig:Fluctuations3D}
\end{figure}

Figure~\ref{fig:Fluctuations3D} shows this data set (open red points) together with the one discussed in the main text (blue points).
The latter dataset was selected for the main text due to improvements in the imaging sequence as reflected in the lower offset variance.
The decrease in both atom number and temperature is due to the evaporation sequence. Figure~3 in the main text corresponds to the projection onto the temperature axis.

Both measurements show the clear signature of the fluctuations, namely the sudden increase as the temperature falls below the critical temperature, followed by a reduction in fluctuations as the atom number and temperature decreases. 
This confirms that the observation of the fluctuations is insensitive to the particular rethermalization conditions.

The observed relative atom number fluctuations at the peak are $ \Delta N_0^2/N = \num{1.6(3)} $ at $ N_\mathrm{max} = \num{2.58(6)e05} $ (blue points) and $ \Delta N_0^2/N = \num{2.9(7)} $ at $ N_\mathrm{max} = \num{4.05(3)e05}$ (red points).  We attribute the larger relative peak fluctuation to the more asymmetric trapping geometry and larger detection uncertainty in the second data set. The fit of Eq.~(2) in the main text to this data set yields $\alpha = 3.0(3)$ and  $ \delta t = -0.015(4)$.

In addition, all data were analysed with a different choice of fitting model for the thermal density. In particular, an extreme model, where $ N_0 $ was taken to be the entire atom number number in the central ROI, was tested. While this overestimates the condensate fraction $ N_0/N $, the defining steep increase in the fluctuations at the critical temperature and subsequent decay was reproduced. This shows that our result is insensitive to the specific choice of fitting model.

\section{Shift of the critical temperature}
A detailed investigation of the mean condensate fraction in comparison with semi-ideal and self-consistent models is beyond the scope of this letter and will be the topic of future work. 
However, this section extends the discussion of the difference between the ideal gas result and the condensate fraction shown in Fig.~2(c) of the main text.

For a finite number of trapped atoms the onset of condensation is a smooth transition, as is evident in Fig.~2(c) of the main text, and a definition of the critical temperature becomes somewhat arbitrary. 
Nonetheless, effects due to the finite size and the interactions are typically discussed in terms of a shift of the critical temperature $ \Delta T_c/T_\mathrm{c}^0 = (T_\mathrm{c} - T_\mathrm{c}^0)/T_\mathrm{c}^0$, where $ T_\mathrm{c} $ is the observed critical temperature and $ T_\mathrm{c}^0 $ the ideal gas value. In general, three effects contribute to the shift, namely the finite size of the system, a mean field modification of the density profile due to interactions and interaction induced correlations. %A positive (negative) shift implies an increase (decrease) of the condensate fraction.

The finite size of the system leads to a reduction of the critical temperature due to the finite zero-point motion of the lowest energy state. The first order contribution is~\cite{Grossmann1995}
\begin{equation}\label{eq:finiteSize}
\frac{\Delta T_\mathrm{c}}{T_\mathrm{c}^0} = -\frac{\zeta(2)}{2\zeta(3)^{2/3}}\frac{ \omega_\mathrm{avg}}{ \bar{\omega}} N^{-1/3},
\end{equation}
where $ \omega_\mathrm{avg} $ and $ \bar{\omega} $ are the arithmetic and geometric averages of the trapping frequencies, respectively. For our experimental parameters at the point of peak fluctuations where $ N_{peak} = \num{2.58(6)e5} $ this corresponds to a shift of $ \Delta T_\mathrm{c}/T_\mathrm{c}^0 = \SI{-1.45}{\percent} $. Additionally, the finite size of the system  leads to a smoothening of the onset of condensation~\cite{Ketterle1996a}.

Interactions modify the critical temperature in two ways~\cite{Holzmann2004}. First, repulsive interactions increase the size of a trapped gas and lower the central density. This also leads to a reduction of the critical temperature and consequently a reduction of the condensate fraction at temperatures below $ T_\mathrm{c}^0 $. Since interactions do not modify the density of a homogeneous gas, this shift is absent in homogeneous systems. The first order shift within mean field theory is given by~\cite{Giorgini1996}
\begin{equation}\label{eq:MeanFieldShift}
\frac{\Delta T_\mathrm{c}}{T_\mathrm{c}^0} = -3.43\frac{a}{\lambda_\mathrm{th}},
\end{equation}
where the ratio of the scattering length to the thermal wavelength is  $ a/\lambda_\mathrm{th} = 0.011$ at the critical temperature in our experiment. This yields a shift of $ \SI{-3.77}{\percent} $. To fully account for the effect of interactions, one must solve the  Hartree-Fock-Popov model self-consistently~\cite{Giorgini1997}. The semi-ideal model provides a convenient simplification by neglecting the repulsion by the thermal atoms, and is compared to experimental data in Fig. 2(c) of the main text.

Finally, interactions induce correlations that cause a positive beyond-mean-field shift of the critical temperature~\cite{Grueter1997,Arnold2001,Kashurnikov2001}. This shift persists in the homogeneous case. Historically, both the magnitude and sign of the shift have been a point of debate~\cite{Holzmann2004}. The beyond-mean-field corrections were experimentally studied for harmonically confined gases~\cite{Smith2011}. This shows that the correction is small when $ a/\lambda_\mathrm{th} \lesssim 0.01$ as in our experiment, in agreement with earlier studies of the condensate fraction~\cite{Gerbier2004}. According to~\cite{Smith2011} the beyond-mean-field correction is estimated to be \SI{0.56}{\percent}.

\section{Classical Field Approximation (CFA)}
The classical field approximation is a method to compute the statistical properties of the quantum many-body system at equilibrium.
In the following we briefly sketch the full quantum description in order to introduce its classical version.

The system at equilibrium at the temperature $T$ is in a mixed state
\begin{equation}
\hat{\rho} = \frac{1}{Z} \,e^{-\beta\hat{H} }
\label{eq:density-matrix}
\end{equation}
where $Z = \mathrm{Tr}\left\{e^{-\beta\hat{H} }\right\}$ is the partition function and $\beta = 1/ (k_{\mathrm B} T)$ is the inverse temperature with $k_{\mathrm B}$ the Boltzmann constant.
We investigate a system of $N$ particles trapped in a symmetric harmonic potential $V(x, y, z) = \frac{1}{2}\,m\, \omega^2 \bb{ x^2+y^2+z^2}$ 
interacting via a contact potential $V_{\mathrm{\twobody}}\bb{\bm{r}_1,\bm{r}_2} = g\,\delta \bb{\bm{r}_1 - \bm{r}_2}$.
The corresponding many body Hamiltonian in second quantization reads
\begin{align}
\hat{H} \, = &\, \int {\mathrm d}^3 r\, \hat{\Psi}^{\dagger} (\bm{r}) \,\hat{h}_1 \, \hat{\Psi} (\bm{r}) \nonumber\\ 
&+ \frac{g}{2}\int {\mathrm d}^3 r\,
\hat{\Psi}^{\dagger} (\bm{r}) \, \hat{\Psi}^{\dagger} (\bm{r}) \hat{\Psi} (\bm{r}) \, \, \hat{\Psi} (\bm{r}),
\label{eq:quantum-ham}
\end{align}
where $\hat{h}_1 = -\frac{\hbar^2 \Delta}{2m} + V(\bm{r})$ is the single particle Hamiltonian and $\hat{\Psi} (\bm{r})$ is the bosonic field operator anihilating a particle at position $\bm{r}$.

Even at ``high'' temperature  $k_{\mathrm B} T\gg \hbar \bb{\omega_x\omega_y\omega_z}^{1/3}$, bosons may accumulate in a single state, the Bose-Einstein condensate. In case of the ideal gas, this orbital is the single particle ground state, i.e. a Gaussian function.  
For interacting atoms, the definition of the condensate is more subtle. Here we use the Penrose-Onsager criterion. In this case, one uses the single particle reduced density matrix
\begin{equation}
\hat{\rho} (\bm{r}, \, \bm{r}') = \frac{1}{N}{\mathrm Tr} \left\{ \hat{\rho}\, \hat{\Psi}^{\dagger}(\bm{r})\hat{\Psi}(\bm{r}')  \right\},
\label{eq:reduced-single-particle-density-matrix}
\end{equation}
which can be uniquely written in the form
\begin{equation}
\hat{\rho} (\bm{r}, \, \bm{r}') = \frac{1}{N}\sum_{l=0}^{\infty} N_l {\phi}^{*}_l(\bm{r})\, {\phi}_l(\bm{r}').
\label{eq:Penrose-Onsager}
\end{equation}
The Penrose-Onsager criterion states that condensation occurs if there exists an eigenvalue $N_l$ that is macroscopic and dominates all other eigenvalues, i.e. $N_l \sim N$. By convention this  dominant eigenvalue is given the index $l=0$. Then the corresponding single-particle wavefunction  $\phi_{l=0} (\bm{r})$ is called the condensate. 

Our goal is to compute the average number of atoms in the condensate $\langle N_{l=0} \rangle$ and the variance of this occupation $\Delta N_0^2 = \langle N_{l=0}^2 \rangle - \langle N_{l=0} \rangle^2$. It is impossible to find these quantities exactly, even numerically. To estimate them we use the classical field approximation.
The central object is the classical field instead of the quantum field operator
\begin{equation}
\hat{\Psi} (\bm{r}) \mapsto \Psi (\bm{r}).
\end{equation}
Then the Hamiltonian operator in Eq.~(\ref{eq:quantum-ham})  becomes the functional
\begin{equation}
\hat{H} \mapsto \mathcal{H}\left[\Psi\right] = 
\int {\mathrm d}^3 r\, {\Psi}^{*} (\bm{r}) \,\hat{h}_1 \, {\Psi} (\bm{r}) + \frac{g}{2}\int {\mathrm d}^3 r\, \, |{\Psi} (\bm{r})|^4
\label{eq:ham-clas}
\end{equation}
The classical field theory of bosons leads to  the ultraviolet catastrophe - a pathologically large occupation of the high momentum modes. The remedy is to appropriately restrict the total phase-space to the low-energy modes only.
We do this by decomposing the classical field $\Psi(\bm{r})$ in the harmonic oscillator basis restricted to the low energy modes
\begin{equation}
\Psi (\bm{r}) = \sum_{i,j, k}^{i+j+k \leq K} \, \alpha_{ijk} h_i(x) h_j(y) h_k(z),
\end{equation}
where coefficients $\alpha_{ijk}$  are complex numbers, $K$ is the cut-off parameter and $h$ denote the eigenstates of the harmonic potential. The cut-off leads to a finite length of the state vector and avoids the ultraviolet catastrophe.
We use a cut-off $K$ that leads to the best approximation of the real probability
distribution. Our cut-off formula for the symmetric trap is
\begin{equation}
K = \left \lfloor{\frac{1}{2} + \frac{T+\mu}{0.33 \bb{T+\mu}^{-1.16}+0.46}}\right \rfloor,
\label{eq:cut-off-formula}
\end{equation}
where $\mu = \mu (N_0, K)$ denotes the chemical potential of the condensate. This formula is a fit to the optimal cut-offs to match the exact results  for (i) average occupation of the condensate of the ideal gas and (ii) the 3D shape of  the condensate at zero-temperature, when the condensate can be obtained simply from the 3D Gross-Pitaevskii equation. At each temperature we determined a self-consistent solution with such number of atoms in the condensate, that its chemical potential $\mu$ matches the chemical potential used in the cut-off Eq.~(\ref{eq:cut-off-formula}).

The Hamiltonian \eqref{eq:ham-clas} and the density matrix become functions of the complex amplitudes $\alpha_{ijk}$, denoted for brevity as the vector $\bm{\alpha}$
\begin{widetext}
	\begin{align}
	\mathcal{H}\bb{\bm{\alpha}} &= \hbar\omega \sum_{i, j, k} \bb{ i +  j +  k } |\alpha_{ijk}|^2 + \int {\mathrm d}^3 r\left|  \sum_{i,j, k}^{i+j+k \leq K} \, \alpha_{ijk} h_i(x) h_j(y) h_k(z) \right|^4 \\
	\mathcal{\rho} \bb{\bm{\alpha}} &= \frac{1}{\mathcal{Z}} e^{-\beta \mathcal{H}\bb{\bm{\alpha}}}\;\delta (N - \sum_{i, j, k} |\alpha_{i,j,k}|^2)\\
	\mathcal{Z} &= \frac{1}{\pi^M}\int {\mathrm d} \bm{\alpha} \,e^{-\beta \mathcal{H}\bb{\bm{\alpha}}}\, \delta (N - \sum_{i, j, k} |\alpha_{i,j,k}|^2),
	\end{align}
\end{widetext}
where $M$ is the number of complex amplitudes. The Dirac deltas in the last two equations express the fact that we work in the canonical ensemble in which the modes' occupations $|\alpha_{ijk}|^2$ sum up to the total number of atoms.

To compute the statistical average of an operator $\hat{O}$ one first has to express it as a function of classical amplitudes $\bm{\alpha}$
\begin{equation}
\hat{O}\mapsto \mathcal{O} \bb{\bm{\alpha}}
\end{equation}
and then compute its average in the phase space of complex amplitudes $\bm{\alpha}$ with the distribution given by $\rho(\bm{\alpha})$
\begin{equation}
\langle \hat{O}\rangle \approx \frac{1}{\pi^M}\int\,{\mathrm d}\bm{\alpha} \,\mathcal{\rho} (\bm{\alpha})\, \mathcal{O} (\bm{\alpha}).
\label{eq:average-classical-field}
\end{equation}
To evaluate the high dimensional integrals we use the Metropolis algorithm.

The first quantity to compute is the reduced density matrix. In the classical field approximation it becomes a finite matrix
\begin{equation}
\hat{\rho} (\bm{r}, \, \bm{r}') \mapsto \left\{ \alpha_{ijk}^{*}\,\alpha_{lmn} \right\}_{ijk  ;\; lmn}.
\end{equation}
Diagonalization of this matrix averaged according to Eq.~(\ref{eq:average-classical-field}) gives the dominant eigenvalue, which is the occupation of the condensate. The corresponding eigenvector $\bm{\alpha}^\mathrm{BEC}$ is the condensate  amplitude. Knowing the condensate vector one can find the second moment of its occupation
\begin{equation}
\langle \hat{N}_0^2 \rangle \approx  \frac{1}{\pi^M}\int\, {\mathrm d}\bm{\alpha}\, \mathcal{\rho}(\bm{\alpha}) \, \left| \bm{\alpha}^{*}\bm{\alpha}^\mathrm{BEC} \right|^4,
\end{equation}
where $\bm{\alpha}^{*}\bm{\alpha}^\mathrm{BEC}$ is the scalar product of the vectors $\bm{\alpha}^{*}$ and $\bm{\alpha}^\mathrm{BEC}$.

The classical field approximation works perfectly in 1D~\cite{Bienias2011, Witkowska2009} and quasi 1D (the Supplementary Material of Ref.~\cite{Karpiuk2012}). It was also successfully used to treat a wide range of phenomena in 2D dipolar system~\cite{Pawlowski2013} and in quasi 2D~\cite{Karpiuk2012,Witkowska2011} (see also~\cite{Brewczyk2007}).

We used the CFA method to estimate the condensate fraction and its fluctuations for the symmetric trap.
The averages given by the integrals in Eq.~(\ref{eq:average-classical-field}) require integrations in spaces with dimensions scaling with the cut-off as $K^3$, which is $O(100)$ even for small cut-offs. With an optimized code running the problem on a large computation grid (Polish computational infrastructure, PL-GRID) one has to restrict the investigation to small total atom numbers and deal with visible statistical errors.
To estimate the role of interactions for the fluctuations we compared the ratio between the fluctuation for the ideal and the interacting gas, both cases computed within CFA. The maximal value of the fluctuations were obtained from a high order polynomial fitted to the numerically computed fluctuations.%central parts of the figures  from the lower panel of Fig. \ref{fig:results-cfa}.

\bibliography{../../references/references_fluctuations_of_a_BEC}